\documentstyle[aps,twocolumn,floats,psfig]{revtex}

\begin{document}

\draft
\title{Coherence properties of the stochastic oscillator}
\author{Martti Havukainen}
\address{Helsinki Institute of Physics, P. O. Box 9, FIN-00014 University of Helsinki, Finland}
\author{Stig Stenholm}
\address{Physics Department, Royal Institute of Technology, Stockholm Sweden}
\date{\today}
\maketitle

\begin{abstract}

An oscillator with stochastic frequency is discussed as a model for
evaluating the quantum coherence properties of a physical system. It is
found that the choice of jump statistics has to be considered with care if
unphysical consequences are to be avoided. We investigate one such model,
evaluate the damping it causes, the decoherence rate and the correlations it
results in and the properties of the state for asymptotically long times.
Also the choice of initial state is discussed and its effect on the time
evolution of the correlations 

\end{abstract}

\section{Introduction}

The harmonic oscillator plays a pivotal role in the development of physics,
both classical and quantum mechanical. It can be solved exactly, it displays a
variety of non-trivial features and it has been applied successfully to a
variety of phenomena. 

A straightforward generalization of the simple oscillator is one with time
dependent frequency; its Hamiltonian is
\begin{equation}
H=\frac{p^2}{2m}+\frac 12m\omega (t)^2x^2.  \label{a1}
\end{equation}
With a deterministic time evolution of  $\omega (t),$ this can describe
parametric frequency generation, radio-frequency traps or  frequency
modulated signal transmission. If, on the other hand, the frequency $\omega
(t)$ is behaving in a random way, the model can be applied to a variety of
physical situations, vide infra. The oscillator with a random frequency, has
been the subject of a variety of theoretical approaches, see e.g. the review
\cite{vanKampen1976}. In quantum theory, the model (\ref{a1}) can be taken as
an approximation to a situation where the oscillator is coupled to some
degrees of freedom carrying out their quantum dynamics in a complicated
fashion. Then it may be possible to model this by a time evolution so
irregular that it can be regarded as stochastic.

With all quantum processes, we are not only interested in the time evolution
of the averaged variables and their noise characterized by the dispersion
from the averages. We also want to know how coherences are conserved, which
variables retain their quantum character and how the classical behavior
emerges. In this paper we undertake the task to investigate these questions
for the oscillator, when $\omega (t)$ is given by a stochastic process.

Stochastic processes can be of two major types: In the one $\omega (t)$
varies continuously according to some model of the Brownian motion type or a
modification of this. The ensuing theory is mathematically well understood,
amenable to analytic approximations and often physically transparent. In
another class of models, random telegraph signals called, the variable $%
\omega (t)$ jumps between constant values according to some stochastic rule.
In simple cases, the consequences of such models can be evaluated \cite{vanKampen1981},
they have simple properties and they can be used as
approximations for continuous behavior in the proper limit \cite{Wodkiewicz1,Wodkiewicz2}.
In a quantum treatment, they offer the advantage that
between the jumps, the well known properties of the harmonic oscillator can
be applied directly. Thus we choose to utilize such a model here, and
evaluate the influence the stochastic evolution has on the quantum coherence
properties.

\section{Physical background}

One application of a stochastic oscillator is to the s.c. continuous
Stern-Gerlach measurement discussed by Dehmelt and his colleagues
\cite{Dehmelt1988,Dehmelt1990,BrownandGabrielse1986}.
Here a trapped particle, an
electron, sees two different trap potentials depending on the internal spin
state. This is changing due to its own dynamics, and by monitoring the
oscillational frequency of the charge, the experimentalist can see the
frequency follow the changes of the internal spin state. This constitutes a
continuous observation of the two quantum states, and the coherence between
them cannot survive for any observable time, and a well defined frequency is
always observed. This continuous Stern-Gerlach effect should be described
by a stochastic quantum model along the lines developed by Carmichael and
colleagues \cite{Carmichael1,Carmichael2}.
They have been able to demonstrate how the
monitoring of the state of a two level system gives precisely the type of
well resolved jumping which is characteristic of the continuous
Stern-Gerlach behavior. We have attempted such modelling, but the
computational capacity needed was found to be unreasonably large, and hence we have
settled on the stochastic process described above. This should suffice when
one wants to investigate the consequences for the survival of quantum
coherence in the oscillator.

Another situation where the time evolution is interrupted by change of the
potential function governing the motion is given by molecular dynamics on
adiabatic energy surfaces. Here spontaneous emission transfers a wave packet
from an excited state to the ground state \cite{LaiandStenholm} or laser
coupling may transfer the state back and forth between the potential levels.
If this process is coherent, Rabi flopping may be observed \cite{PaloviitaandSuominen}
or adiabatic motion may ensue \cite{GarrawayandSuominen}. With less
coherent coupling, the dynamic evolution has been described by surface
hopping \cite{Tully}, where the deterministic motion on one potential surface
is stochastically transferred to another one. It is well justified to
describe this process by a random switch between the various potential
curves. In this paper, the model assumes that the two harmonic states are
situated directly above each other; this may not be a realistic description
of a molecule, but it simplifies the treatment, and does not greatly affect
the features we are investigating. It would not, in principle, be difficult
to carry out the simulations on more realistic modifications of this model.

When an evolving quantum state is suddenly transferred to another potential
at random times, we expect the quantum coherence to be obliterated, and the
system to behave in a more classical way. This is seen in the density matrix
in such a way that it tends towards the diagonal both in the position and
the momentum representation. The uncertainties along the diagonal are then
large, their product is expected to greatly exceed the minimum allowed by
quantum theory. We have performed calculations to observe such disappearence
of coherence.

The simplest model is to let the stochastic frequency jump between two
values $\omega _1$ and $\omega _2$ with the constant jump probability $\nu .$
This model will be found to have physically less satisfactory features: The
coherences decay, but the jumping process feeds energy into the system so
efficiently that the energy grows at a rate quite comparable to the rate of
vanishing of the coherences. The explanation for this strange behavior is in
the physics of the oscillator motion. The system spends much of its time
near the classical turning points of the oscillation, and consequently the
jumps most likely take place here. Whenever the state jumps from the turning
point of the flatter (slower) potential to the steeper (faster) one, it
appears far up the potential slope, and a large increase in energy ensues.
Jumping in the opposite direction, the system looses energy, but not enough
to compensate for the gain. Thus the rapid increase in energy.

To overcome the unphysical features of our model, we strive to conserve
energy in the jump process. This means that we have to favor jumping in the
region where the change of potential energy is minimal, i.e. near the bottom
of the wells. Thus the wave function should be most eager to jump, when it
overlaps most with the ground state; then only the kinetic energy is
transferred between the levels. This also maximizes the conservation of
linear momentum; near the potential minimum, the motion most resembles free
propagation.

In order to achieve the goal outlined above, we choose the following model:
When the state on level $1$ is $\Psi _1(t)$, we jump to level $2$ with the
probability
\begin{equation}
P_{12}(t)=\nu \mid \langle \Psi _2^0\mid \Psi _1(t)\rangle \mid ^2,
\label{a2}
\end{equation}
where $\Psi _2^0$ is the ground state of level $2.$ The jumps are taken to
occur at a random time determined by $P_{12}$, and after that the state $%
\Psi _1(t)$ is transferred to level $2$ and evolved according to the
corresponding potential. The probability to jump back to level $1$ is now
determined by $P_{21}(t)$ given by the expression (\ref{a2}) with $1$
and $2$ interchanged. This simulation continues, and enough numerical data
are accumulated to allow us to evaluate the ensemble averaged quantities.
The results are then compared with those obtained by the simple model having 
$P=\nu $ constant.

\section{Time evolution of the state}

The Schr\"{o}dinger time evolution with the potential (\ref{a1}) is given by
\begin{equation}
i\hbar \frac \partial {\partial t}\Psi (x,t)=-\frac{\hbar ^2}{2m}\frac{%
\partial ^2\Psi (x,t)}{\partial x^2}+\frac 12m\omega _i^2x^2\Psi (x,t),
\label{a3}
\end{equation}
where $i=1$ or $2.$ Using the standard definition of the Wigner function
\cite{Hilleryetal1984} 
\begin{equation}
W(X,P)=\frac 1{2\pi \hbar }\int dx\;\langle X+\frac x2\mid \hat{\rho} \mid
X-\frac x2\rangle \exp \left( -i\frac{Px}{\hbar }\right) ,  \label{a4}
\end{equation}
we can obtain the equation of motion
\begin{equation}
\frac{\partial W(X,P)}{\partial t}+\left( \frac Pm\right) \frac{\partial
W(X,P)}{\partial X}-m\omega _i^2X\;\frac{\partial W(X,P)}{\partial P}=0.
\label{a5}
\end{equation}
For the harmonic oscillator, this can be solved, and the various moments
evaluated. It is, however, easy to see that the equations of motion for the
moments close in each order, and their evolution equations can be obtained
by simple calculations from that of the Wigner function. The mixed moments
of the type $\langle xp\rangle $ denote symmetrized products, as these are
given by the Wigner function.

The equations for the first moments we find to be the classical dynamic
equations
\begin{equation}
\frac d{dt}\left( 
\begin{array}{r}
\langle x(t)\rangle  \\ 
\\ 
\langle p(t)\rangle 
\end{array}
\right) =\left( 
\begin{array}{ccc}
0 &  & \frac 1m \\ 
&  &  \\ 
-m\omega ^2 &  & 0
\end{array}
\right) \left( 
\begin{array}{r}
\langle x(t)\rangle  \\ 
\\ 
\langle p(t)\rangle 
\end{array}
\right) .  \label{a6}
\label{firstmoments}
\end{equation}
For the second order ones we find
\begin{eqnarray}
\label{secondmoments}
\lefteqn{\frac{d}{dt}\left(
\begin{array}{c}
\langle x^2(t)\rangle  \\ 
\langle p^2(t)\rangle  \\ 
\langle x(t)p(t)\rangle 
\end{array}
\right) =} \nonumber \\
& & \left(
\begin{array}{ccccc}
0 & 0 & \frac 2m \\ 
0 & 0 & -2m\omega ^2 \\ 
-m\omega ^2 &  \frac 1m & 0
\end{array}
\right)
\left( 
\begin{array}{c}
\langle x^2(t)\rangle  \\ 
\langle p^2(t)\rangle  \\ 
\langle x(t)p(t)\rangle 
\end{array}
\right) 
\end{eqnarray}
Given the initial values, it is straightforward to integrate these equations
for the moments. From these we can calculate the variances
\begin{eqnarray}
\sigma _x^2(t) & = & \langle x^2(t)\rangle -\langle x(t)\rangle ^2 \\ 
\sigma _p^2(t) & = & \langle p^2(t)\rangle -\langle p(t)\rangle ^2 \\ 
\sigma _{xp}^2(t) & = & \langle x(t)p(t)\rangle -\langle x(t)\rangle \langle
p(t)\rangle 
.  \label{a8}
\end{eqnarray}
If the initial quantum state is a Gaussian wave function, it suffices to
integrate the moments to the first two orders; it is known that the state
will stay Gaussian, and that it is described by its moments of the first two
orders. A straightforward calculation shows that the Schr\"{o}dinger
equation is satisfied by the state
\begin{eqnarray}
\Psi (x,t) = (2\pi\sigma_x^2(t))^{-1/4}\exp ( -\frac{(x-\langle x(t)\rangle)^2}{4\sigma_x^2(t)}+ \nonumber \\
\frac{i\sigma _{xp}^2(t)}{2\hbar \sigma _x^2(t)}(x-\langle x(t)\rangle)^2 +\frac{i\langle p(t)\rangle }{\hbar }(x-\langle x(t)\rangle) +i\theta (t) )  ,
\label{a9}
\end{eqnarray}
where the phase $\theta (t)$ disappears in the calculation of the Wigner
function (\ref{a4}) and thus does not affect the physics. All moments
occurring in the state are not independent. If we start from a minimum
uncertainty state, the combination
\begin{equation}
\sigma _x^2(t)\sigma _p^2(t)-\left( \sigma _{xp}^2(t)\right) ^2=\frac{%
\hbar ^2}4.  \label{a10}
\end{equation}
is conserved.

The density matrix in the position representation is given by $\langle
x_1\mid \hat{\rho} \mid x_2\rangle ,$ and its quantum coherence is the amount of
off-diagonality, i.e. the dependence on the variable $x=x_1-x_2.$ Its moments can be evaluated
directly from the Wigner function according to
\begin{eqnarray}
\langle x^n\rangle  & = & \int \int dx_1dx_2\left( x_1-x_2\right) ^n\langle
x_1\mid \hat{\rho} \mid x_2\rangle  \nonumber \\ 
& = & {2\pi
(i\hbar )^n\int dR\left[ \frac{\partial ^n}{\partial P^n}W(R,P)\right]
_{P=0}.}
\label{a11}
\end{eqnarray}
A similar result can be obtained for the off-diagonality in the $p$%
-representation.

\section{Numerical results}

Starting from a Gaussian initial state, the equations (\ref{firstmoments}) and (\ref
{secondmoments}) are integrated, and the state can be directly obtained from Eq.(\ref
{a9}). We have found this integration to be orders of magnitude faster than
the direct integration of Schr\"{o}dinger's equation, and obtaining the
state from (\ref{a9}), we can directly calculate the Wigner function. The
most time consuming part is still to accumulate enough single histories to
obtain the ensemble averaged Wigner function and the corresponding average
moments. This computation is, however, straightforward, and consists in
simulating histories with the proper jump probabilities given as explained
above. The whole calculation is time consuming but essentially trivial.

In the numerical work we want to evaluate the average position and momentum,
and their variances. We have used dimensionless scaled units, which bring
the numerical results into ranges of order unity. We have integrated the
above problem using a Gaussian initial state on level $1$ with the initial
values
\begin{equation}
\begin{array}{rrrrr}
\langle x(0)\rangle =2.0 & ; & \langle p(0)\rangle =0 & ; &  \\ 
\sigma _x^2(0)=0.5 & ; & \sigma _p^2(0)=0.5 & ; & \sigma _{xp}^2(0)=0.
\end{array}
\label{a12}
\end{equation}
The stochastic model consists of jumping between the two frequencies $\omega
_1=0.7$ and $\omega _2=1.2$ with the jump frequency parameter $\nu =0.8.$ We
use two models, in the one the jump probabilities are constant
\begin{equation}
P_{12}=P_{21}=\nu ,  \label{a13}
\end{equation}
in the other one we use the nonconstant, time dependent jump probabilities
determined as explained in Sec.2 and Eq.(\ref{a2}). In order to get
satisfactory statistics for evaluating the ensemble averages, the number of
individual histories making up the ensemble consists of $N=30000$ runs.

In Fig. 1 we report the energy of the oscillator as function of
dimensionless time. The solid line is the constant jump probability result,
which displays a nearly exponential growth. The dotted line is the result
with the nonconstant jump probability determined according to Eq.(\ref{a2}).
We can see that the goal declared has been achieved; the energy grows much
more slowly and clearly not exponentially. This is ascribed to the tendency
of the nonconstant jump model to concentrate the jumping to the center of
the potential.

\begin{figure}[htp]
\vspace{-1.0cm}
\centerline{\psfig{file=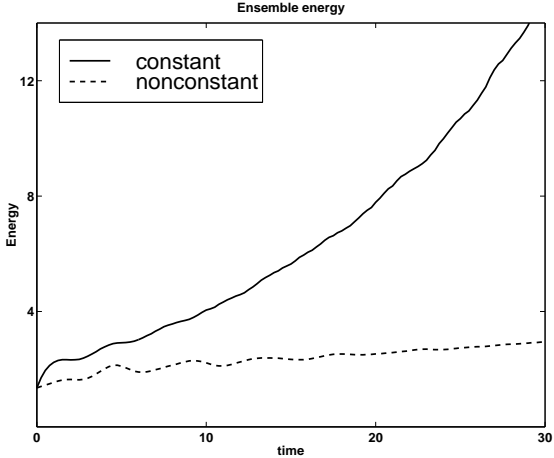,width=8.5cm,angle=90,bbllx=0.0cm,bblly=0.0cm,bburx=22cm,bbury=26cm,clip=}}
\vspace{-0.5cm}
\caption{The ensemble energy as a function of time with constant and nonconstant jump
probability. The jump frequency is $\nu=0.8$. The initial wavefunction is Gaussian with
parameters $\langle X(0)\rangle = 2.0$, $\langle P(0)\rangle = 0.0$,
$\langle \sigma_X^2(0)\rangle = 0.5$, $\langle \sigma_P^2(0)\rangle = 0.5$,
$\langle \sigma_{XP}^2(0)\rangle = 0.0$. The frequencies are $\omega_1=0.7$ and $\omega_2=1.2$.
The initial frequency is $\omega_1$. The ensemble used is of size $N=30000$.}
\label{E}
\end{figure}

That this is, indeed, the case can be seen from Figs. 2 and
3. In Fig. 2 we show the statistical distribution of jumps from the level $1$
to $2.$ The classical turning points of the oscillational motion are
indicated by the solid bars on the horizontal axis. The solid line shows the
result for the constant probability, which leads to many jumps near or
outside the turning points, where the oscillator spends large times. The
peak at the right hand edge indicates the position of the starting wave
packet. The dashed line is the result of the simulation with nonconstant
probabilities. The jumps occur much closer to the center of the potentials
as expected, even if the effect of the turning points is still seen. The
total jump rate has decreased, thus the initial asymmetry is no longer so
manifest. 
\begin{figure}[htp]
\vspace{-4.0cm}
\hspace{0.2cm}\centerline{\psfig{file=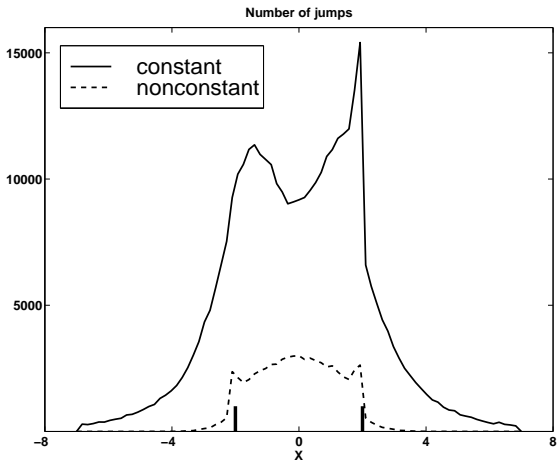,width=8.5cm,angle=90,bbllx=0.0cm,bblly=0.0cm,bburx=22cm,bbury=26cm,clip=}}
\vspace{-0.5cm}
\caption{The number of jumps from the potential with $\omega_1$ to the potential with $\omega_2$.
The bars on the $X$-axis show the classical turning points with the frequency
$\omega_1$. The parameters are the same as in FIG. 1}
\label{jumpvec1to2}
\end{figure}
In Fig. 3, the distribution of jumps from state  $2$ to state 1 is
shown. The constant probability case shown by the solid line still
concentrate to outside the turning points in the steeper potential; these
are again shown as solid bars. The symmetry is now in the opposite direction
because, with the initial condition chosen,  there must occur one initial
jump to state $2$ before we can have a jump back. The dashed line shows the
result for the model with nonconstant probabilities. The same features are
seen as in the previous figure, but in this case of faster oscillations, the
turning points have slightly enhanced effect. 
\begin{figure}[htp]
\vspace{0.0cm}
\hspace{0.2cm}\centerline{\psfig{file=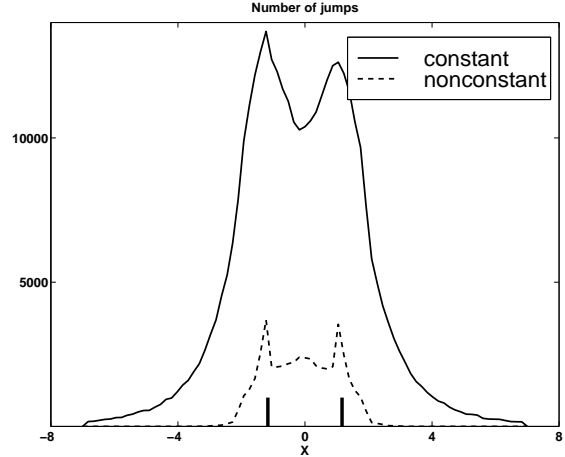,width=8.5cm,angle=90,bbllx=0.0cm,bblly=0.0cm,bburx=22cm,bbury=26cm,clip=}}
\vspace{-0.5cm}
\caption{The number of jumps from the potential $\omega_2$ to $\omega_1$. The parameters are the same as in FIG. 1}
\label{jumpvec2to1}
\end{figure}
The decrease of energy growth in the case of nonconstant jump probabilities
allows the model to damp the average motion faster. This is seen in both the
position variable, Fig. 4, and the momentum variable, Fig. 5. These variables are
expected to behave in very similar manner for the harmonic motion. The
damping time can be estimated to be of the order of $10$ units, which is
much less than the one observed with constant probability.
\begin{figure}[htp]
\vspace{0.0cm}
\hspace{0.2cm}\centerline{\psfig{file=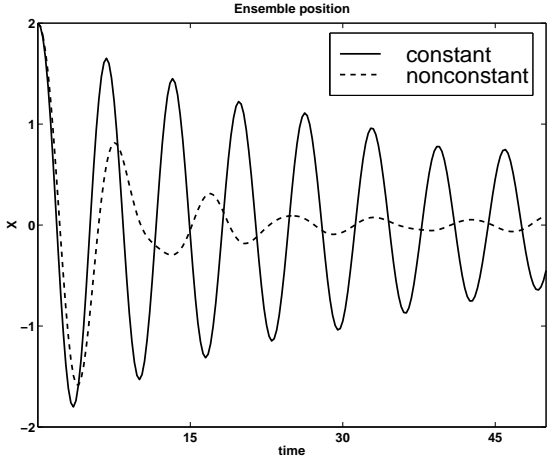,width=8.5cm,angle=90,bbllx=0.0cm,bblly=0.0cm,bburx=22cm,bbury=26cm,clip=}}
\vspace{-0.5cm}
\caption{Ensemble position as a function of time. The parameters are the same as in FIG. 1}
\label{X}
\end{figure}
\begin{figure}[htp]
\vspace{0.0cm}
\hspace{0.2cm}\centerline{\psfig{file=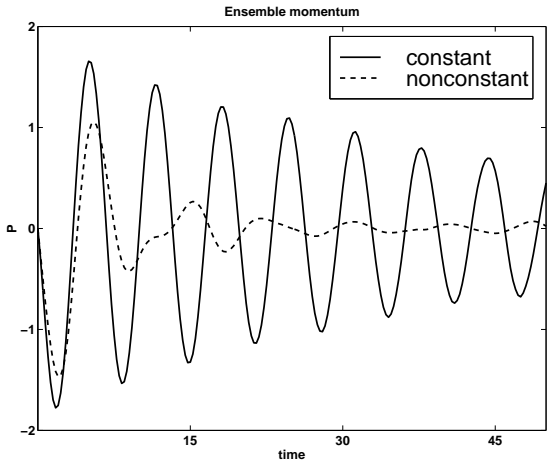,width=8.5cm,angle=90,bbllx=0.0cm,bblly=0.0cm,bburx=22cm,bbury=26cm,clip=}}
\vspace{-0.5cm}
\caption{Ensemble momentum as a function of time. The parameters are the same as in FIG. 1}
\label{P}
\end{figure}
For harmonic
motion, the second moments of both position and momentum are expected to
grow like the energy, see Fig. 1. This is also seen in the variances; for
constant probabilities they increase exponentially, while the nonconstant
case grows more slowly. This is shown in Fig. 6, which should be compared
with Fig. 1; the momentum variance behaves in a very similar manner.
\begin{figure}[htp]
\vspace{0.5cm}
\hspace{0.2cm}\centerline{\psfig{file=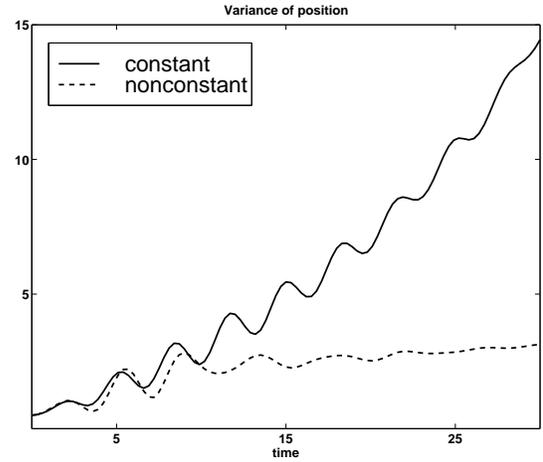,width=8.5cm,angle=90,bbllx=0.0cm,bblly=0.0cm,bburx=22cm,bbury=26cm,clip=}}
\vspace{-0.5cm}
\caption{Position variance as a function of time. The parameters are the same as in FIG. 1}
\label{varX}
\end{figure}
For the model with nonconstant jump probabilities, the time scale of
motional damping has been found to be of order $10$, the range over which
energy grows is several times this value, see Fig. 1. If we look at the
dephasing causing the disappearance of quantum correlations in the position
variable, we plot the expectation value
\begin{equation}
\langle x^2\rangle =\int \int dx_1dx_2\left( x_1-x_2\right) ^2\langle
x_1\mid \hat{\rho} \mid x_2\rangle ,  \label{a14}
\end{equation}
which can be obtained directly from the Wigner function as shown in Eq. (\ref
{a11}). The position density matrix can also be obtained from the Wigner
function by inverting the relation (\ref{a4})
\begin{eqnarray}
\langle x_1\mid \hat{\rho} \mid x_2\rangle =\int dP\;\exp \left( \frac{iP(x_1-x_2)%
}{\hbar }\right)\cdot \nonumber\\
\hspace{1cm} W\left( \frac{(x_1+x_2)}2,P\right) .  \label{a15}
\end{eqnarray}
Fig. 7 shows the decay of the quantum mechanical correlations, and we can
see that they disappear after a time of approximately $4$ units, which is
considerably faster than the other time scales. In particular, the increase
in energy is totally negligible over this time for the nonconstant jump
probability; see Fig. 1. This clearly corrects the unphysical behavior of the
model with constant probability. The constant model, on the other hand, destroys the
quantum coherence about as fast as the other model does, but the asymptotic
value for large times does not seem to vanish but linger on at a small but
finite value. However, the very process of jumping removes the possibility
to retain coherence, the actual statistics of the process seems to matter
less.
\begin{figure}[htp]
\vspace{0.5cm}
\hspace{0.2cm}\centerline{\psfig{file=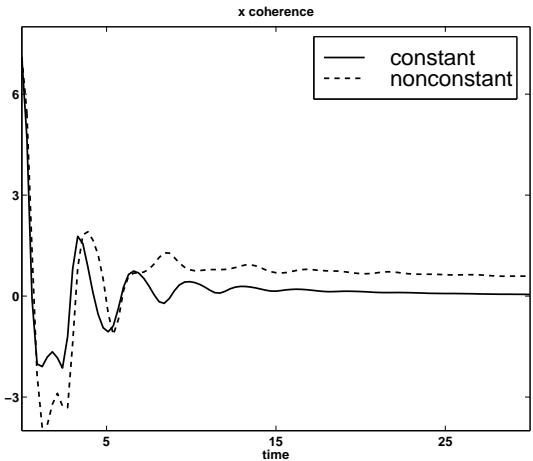,width=8.5cm,angle=90,bbllx=0.0cm,bblly=0.0cm,bburx=22cm,bbury=26cm,clip=}}
\vspace{-0.5cm}
\caption{$X$-coherence as a function of time. The parameters are the same as in FIG. 1}
\label{x}
\end{figure}

When the harmonic oscillator is subjected to a random perturbation, we
expect the final result to resemble a thermal distribution. In order to
check this, we plot the density matrix in the occupation number
representation. This is directly obtained from the position representation (%
\ref{a15}) using the relation
\begin{equation}
\langle n_1\mid \hat{\rho} \mid n_2\rangle =\int \int dx_1dx_2\langle n_1\mid
x_1\rangle \langle x_1\mid \hat{\rho} \mid x_2\rangle \langle x_2\mid n_2\rangle .
\label{a16}
\end{equation}
The states $\langle x\mid n\rangle $ are the Hermit polynomial eigenstates
of the harmonic oscillator, and the integrations in Eq.(\ref{a16}) are thus
straightforward.

The diagonal elements of the density matrix in the occupation number
representation with the frequency $\omega_1$ are shown in Fig. 8 for
both models at the asymptotically
large time $30$. We find that the model with nonconstant probability
decreases much faster and goes to zero for large values of the quantum number.
Both follow asymptotically an exponential decrease with the quantum number $n$.
\begin{figure}[htp]
\vspace{0.5cm}
\hspace{0.2cm}\centerline{\psfig{file=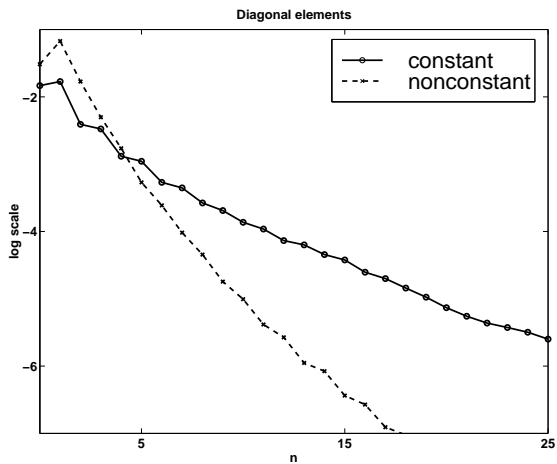,width=8.5cm,angle=90,bbllx=0.0cm,bblly=0.0cm,bburx=22cm,bbury=26cm,clip=}}
\vspace{-0.5cm}
\caption{Diagonal elements of the ensemble energy density matrix $\langle m|\hat{\varrho}|n\rangle$
as a function of time. The Fock-states have the frequency $\omega_1$. The scale on the y-axis is logarithmic. The parameters are the same as in FIG. 1}
\end{figure}

The exponential decrease indicates that the state is close to a thermal
distribution of the Planck type. However, the states near $n=0$ clearly
deviate more than  the following ones from a
thermal distribution. This is verified when we follow the time evolution of
the Wigner function in the $X,P-$plane. In Fig. 9, the original Gaussian
state at $t=0$, rotates and distorts due to the influence of the jumping,
which here is taken with nonconstant probabilities. The fore front
rotates faster and the distribution loses phase information within a few
units of the time. In Fig. 10, the time variable is $30$, and the phase
information has disappeared. This picture corresponds to the same situation as that in
Fig. 8, and the dip seen in the middle corresponds to the minimum at $n=0$
in that figure. It is obvious, that, in the occupation number
representation, there are no off-diagonal density matrix elements in the
state shown in Fig. 10.
\begin{figure}[htp]
\vspace{0.0cm}
\hspace{0.2cm}\centerline{\psfig{file=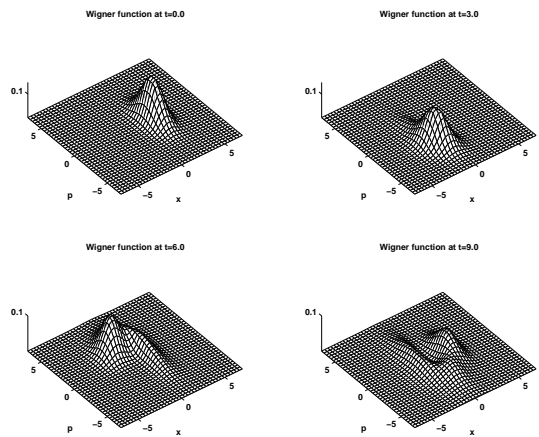,width=8.5cm,angle=90,bbllx=0.0cm,bblly=0.0cm,bburx=22cm,bbury=26cm,clip=}}
\vspace{-0.5cm}
\caption{The Wigner function at four different times with a nonconstant jump probability. The parameters are the same as in FIG. 1}
\label{meshwigner4figs}

\end{figure}
\begin{figure}[htp]
\vspace{0.5cm}
\hspace{0.2cm}\centerline{\psfig{file=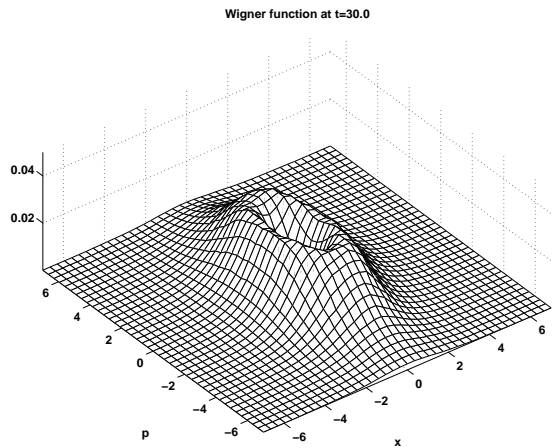,width=8.5cm,angle=90,bbllx=0.0cm,bblly=0.0cm,bburx=22cm,bbury=26cm,clip=}}
\vspace{-0.5cm}
\caption{The Wigner function at $t=30.0$ with a nonconstant jump probability. The parameters are the same as in FIG. 1}
\label{meshwignerfinal}
\end{figure}

\section{Effect of initial squeezing}

The numerical results obtained so far have started from the ideal minimum
uncertainty state (\ref{a12}). In order to see how sensitive our results are
to this assumption, we have integrated the equations also with squeezed
initial states. In the X-squeezed state we have set $\sigma _X^2(0)=0.25$
and $\sigma _P^2(0)=1.0$ ; in the P-squeezed state we have $\sigma
_X^2(0)=1.0$ and $\sigma _P^2(0)=0.25.$ Fig. 11 reports the growth of energy
in the model with nonconstant jump probabilities; compare this figure with
Fig. 1. For the P-squeezed state, the advantages of letting the jump
probability vary is partly lost; the X-squeezed case is not very different
from that shown in Fig. 1. The wave packet still spends much time near the
turning points, and in the case of P-squeezing at the initial point, the
wave packet is very broad here. This increases the overlap with the central
region of the oscillators, and the jumps adding energy to the system become
more numerous.
\begin{figure}[htp]
\vspace{0.5cm}
\hspace{0.2cm}\centerline{\psfig{file=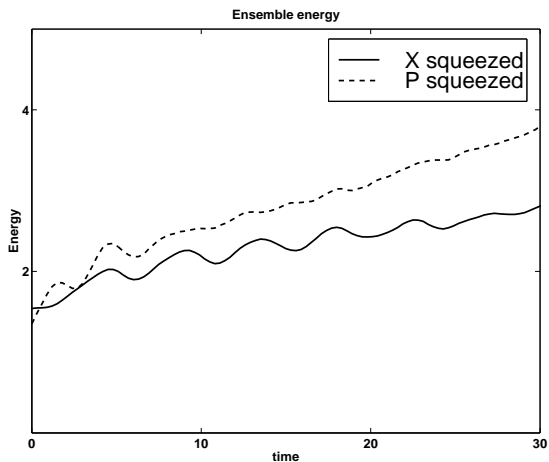,width=8.5cm,angle=90,bbllx=0.0cm,bblly=0.0cm,bburx=22cm,bbury=26cm,clip=}}
\vspace{-0.5cm}
\caption{
The ensemble energy as a function of time with X-squeezed $\sigma_X^2(0) = 0.25$,
$\sigma_P^2(0) = 1.0$ and P-squeezed $\sigma_X^2(0) = 1.0$, $\sigma_P^2(0) = 0.25$ as initial
conditions. The other parameters are the same as in FIG 1.}
\label{Esq}
\end{figure}
In Fig. 12 we show the jump distribution from the initial level 
$1$ to level 2, and we see that the P-squeezed state clearly tends to
concentrate around the turning points. This figure should be compared with
Fig. 2, and we can see that even the X-squeezed state has an enhanced jump
probability as compared with the unsqeezed result, the dashed line in Fig. 2.
Fig. 12, does not, however, show any asymmetry as found in the model with a
constant probability. The jump frequency from level $2$ back to level $1$
shows a behavior almost identical to that of Fig. 12; the P-squeezed state
tends to jump around the turning points, the distribution has a minimum at
the center, which is not found in the case of X-squeezing. 
\begin{figure}[htp]
\vspace{0.5cm}
\hspace{0.2cm}\centerline{\psfig{file=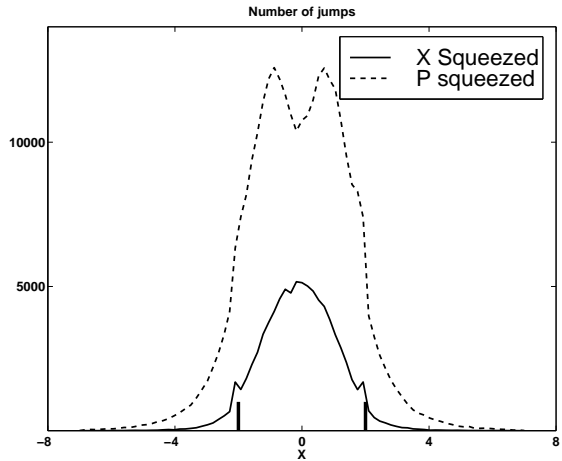,width=8.5cm,angle=90,bbllx=0.0cm,bblly=0.0cm,bburx=22cm,bbury=26cm,clip=}}
\vspace{-0.5cm}
\caption{The number of jumps from the potential with $\omega_1$ to the potential with $\omega_2$.
The bars on the x-axis show the classical turning points with the frequency
$\omega_1$. The parameters are the same as in FIG. 12}
\label{jumpvec1to2sq}
\end{figure}
Fig. 13 shows how the average position is decaying. There is little
difference between the P- and X-squeezed cases, and only a slight decrease
in the decay rate is seen when compared with the decay in Fig. 4. The average
momentum behaves in a very similar way.
\begin{figure}[htp]
\vspace{0.5cm}
\hspace{0.2cm}\centerline{\psfig{file=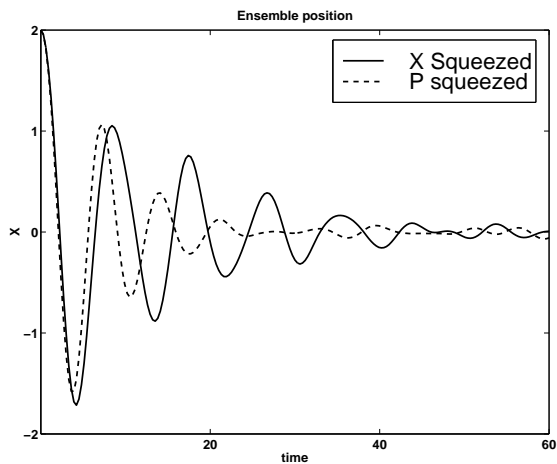,width=8.5cm,angle=90,bbllx=0.0cm,bblly=0.0cm,bburx=22cm,bbury=26cm,clip=}}
\vspace{-0.5cm}
\caption{Ensemble position as a function of time. The parameters are the same as in FIG. 12}
\label{Xsq}
\end{figure}

The variance in the position is
shown in Fig. 14. For the case of X-squeezing, the wave packet is broad near
the center, where the jumps take place, and the jumping consequently adds
little width to the packet; the behavior is very similar to that of the
dashed curve in Fig. 6 showing the result for the unsqueezed case. The
P-squeezed case, however, jumps more often near the turning points, and
consequently the width in position space is growing more rapidly. However,
not by far as rapidly as with the constant jump probability shown in Fig. 6,
the solid curve.
\begin{figure}[htp]
\vspace{0.5cm}
\hspace{0.2cm}\centerline{\psfig{file=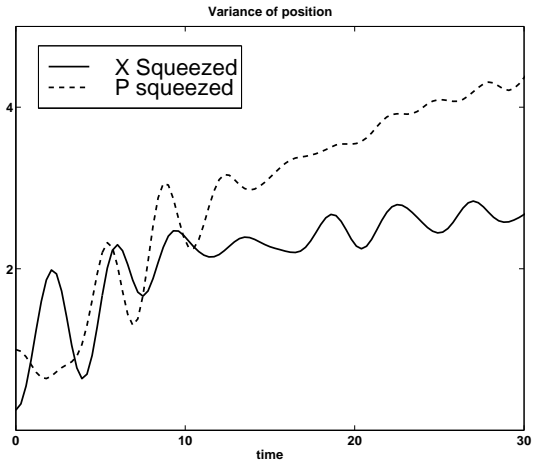,width=8.5cm,angle=90,bbllx=0.0cm,bblly=0.0cm,bburx=22cm,bbury=26cm,clip=}}
\vspace{-0.5cm}
\caption{Position variance as a function of time. The parameters are the same as in FIG. 12}
\label{varXsq}
\end{figure}
\begin{figure}[htp]
\vspace{0.5cm}
\hspace{0.2cm}\centerline{\psfig{file=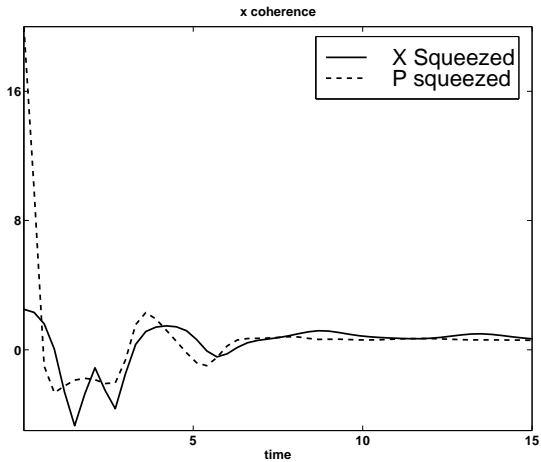,width=8.5cm,angle=90,bbllx=0.0cm,bblly=0.0cm,bburx=22cm,bbury=26cm,clip=}}
\vspace{-0.5cm}
\caption{$X$-coherence as a function of time. The parameters are the same as in FIG. 12}
\label{xcoherencesq}
\end{figure}
 When we look at the disappearance of the quantum coherence, the
variable (\ref{a14}), the result in Fig. 15 shows that squeezing does not
affect the rate of disappearance of quantum coherence much as compared with
Fig. 7. Jumping always destroys the coherence rapidly and efficiently as
expected.

\section{Conclusion }

In this paper we have discussed the time evolution of quantum coherence and
correlations in a harmonic oscillator with a stochastic frequency.
For simplicity we choose a jump process, where the frequency stays constant
but switches randomly according to a stochastic rule. In Sec.II of the paper
we discuss the possible physical situations which may be modelled by such
behaviour. 

We find that one should be careful in modelling random jumping, because the
enforced transitions violate energy conservation, which leads to clearly
unphysical growth of the oscillator energy. In order to amend this
shortcoming, we device an ad hoc model, where jumping near the potential
minimum is enhanced, which tends to conserve energy. In addition, the motion
resembles free propagation most closely at this point, and hence the jump
achieves momentum conservation as close as possible, when we jump here. This
gives some additional physical motivation to our model, when we consider it
as an attempt to emulate the behaviour of real physical processes.

The model suggested is found to achieve the asserted goal: the energy grows
but little, and the decoherence time scale is much shorter than that over
which the energy changes. We find that jumping in itself tends to destroy
quantum coherence, but jumping near the center of the potentials achieves
this without a build-up of unphysical energy. After an
asymptotically rather long time, the diagonal elements of the density matrix
resembles that of a thermal state, but the occurrence of very low quantum
numbers is less likely than it would be thermally. This is verified by
plotting the Wigner function, which displays a ring shaped form resembling
that of an operating laser. No phase information survives for large times.

We have always started the integration near one of the turning points.
Squeezing the state in the P-direction, we find that the accompanying
broadening in the X-direction leads to more jumps near the turning points.
This trends to counteract the improvements introduced by our nonconstant
jump probability model. With squeezing in the X-direction less ill effects
are found. In this case, they would, however, be seen if we started the wave
packet closer to the center of the potial. 

We conclude that the model of a stochastic oscillator can be used to
investigate the quantum coherences, but care has to be excersised when the
model is chosen. In our work, we have shown how to achieve physically
reasonable results with a suitably modified jump rate. Which model is
appropriate in a given physical system must be left open; different
situations may require different choices of jump statistics.

\end{document}